\documentclass{an}
\usepackage{times}
\usepackage{graphicx}
\graphicspath{{./fig/}{./png/}}
\usepackage{fancyhdr}
\usepackage{bm}
\newcommand{\EQ}{\begin{equation}}
\newcommand{\EN}{\end{equation}}
\newcommand{\EQA}{\begin{eqnarray}}
\newcommand{\ENA}{\end{eqnarray}}
\newcommand{\eq}[1]{(\ref{#1})}
\newcommand{\EEq}[1]{Equation~(\ref{#1})}
\newcommand{\Eq}[1]{Eq.~(\ref{#1})}

\newcommand{\Fig}[1]{Fig.~\ref{#1}}

\newcommand{\Tab}[1]{Table~\ref{#1}}
\newcommand{\Figs}[2]{Figs~\ref{#1} and \ref{#2}}

\newcommand{\bra}[1]{\langle #1\rangle}

\newcommand{\meanEMF}{\overline{\vec{\cal E}}}

\newcommand{\meanB}{\overline{B}}

\newcommand{\meanBB}{\overline{\vec{B}}}
\newcommand{\meanJJ}{\overline{\vec{J}}}

\newcommand{\meanWW}{\overline{\vec{W}}}

\newcommand{\uu}{{\vec{u}}}
\newcommand{\BB}{{\vec{B}}}

\newcommand{\jj}{{\vec{j}}}

\newcommand{\bb}{{\vec{b}}}

\newcommand{\nab}{\mbox{\boldmath $\nabla$} {}}

\newcommand{\oo}{\mbox{\boldmath $\omega$} {}}

\def\onethird{{\textstyle{1\over3}}}

\newcommand{\yapj}[3]{ #1, {ApJ,} {#2}, #3}

\newcommand{\yan}[3]{ #1, {AN,} {#2}, #3}

\newcommand{\yana}[3]{ #1, {A\&A,} {#2}, #3}

\newcommand{\ygafd}[3]{ #1, {GApFD,} {#2}, #3}
\newcommand{\yjfm}[3]{ #1, {JFM,} {#2}, #3}
\newcommand{\ypf}[3]{ #1, {PhFl,} {#2}, #3}

\newcommand{\yprl}[3]{ #1, {PRL,} {#2}, #3}
\newcommand{\ypre}[3]{ #1, {PRE,} {#2}, #3}

\newcommand{\ymn}[3]{ #1, {MNRAS,} {#2}, #3}

\newcommand{\yjour}[4]{ #1, {#2}, {#3}, #4}

\begin{document}

\title{Simulations of the anisotropic kinetic and magnetic alpha effects}
\author{A.\ Brandenburg\inst{1} \and K.\ Subramanian\inst{2}}

\institute{
NORDITA, Roslagstullsbacken 23,
AlbaNova University Center, 106 91 Stockholm, Sweden
\and
IUCAA, Post Bag 4, Pune University Campus, Ganeshkhind, Pune 411 007, India
}

\date{Received 2007 March 19, accepted 2007 April 2}

\abstract{
Using simulations of isotropically forced helical turbulence the
contributions to kinetic and magnetic alpha effects are computed.
It is shown that for the parameter regimes considered in an
earlier publication (Brandenburg \& Subramanian 2005, A\&A 439, 835),
the expressions for isotropic and anisotropic alpha effects give
quantitatively similar results.
Both kinetic and magnetic alpha effects are proportional to a
relaxation time whose value, in units of the turnover time, is shown to
be approximately unity and independent of the magnetic Reynolds number.
}

\maketitle

\section{Introduction}

In a seminal paper, Pouquet al al.\ (1976) showed that in the nonlinear
regime the alpha effect in mean field magnetohydrodynamics is no
longer governed by the kinetic helicity (Steenbeck et al.\ 1966), but
there is an additional contribution from the current helicity, so
\EQ
\alpha=\onethird\tau
\left(-\overline{\oo\cdot\uu}
+\rho_0^{-1}\overline{\jj\cdot\bb}\right),
\label{alpiso}
\EN
where $\tau$ is a correlation or relaxation time,
$\uu$ is the small scale velocity,
$\oo=\nab\times\uu$ is the small scale vorticity,
$\bb$ is the small magnetic field, and
$\jj=\nab\times\bb/\mu_0$ is the small scale current density.
Overbars denote some suitable form of averaging.
\EEq{alpiso} has been used to explain cata\-stro\-phic
(magnetic Reynolds number dependent) quenching of the
alpha effect in the nonlinear regime (Gruzinov \& Diamond 1994;
Bhattacharjee \& Yuan 1995; Field \& Blackman 2002).
Technically, the $\overline{\jj\cdot\bb}$ term arises naturally
when the $\tau$ approximation is used (Kleeorin \& Rogachevskii 1999;
R\"adler et al.\ 2003; Blackman \& Field 2002, 2003; see review by
Brandenburg \& Subramanian 2005a).

In a recent paper, Brandenburg \& Subramanian (2005b, hereafter BS05)
presented results of numerical simulations that demonstrate the
rise of the $\overline{\jj\cdot\bb}$ term with magnetic Reynolds number
in the presence of a finite imposed magnetic field, $\meanBB_0$.
Recently, R\"adler \& Rheinhardt (2007) have pointed out that for finite
values of $\meanB_0$ it may be important to consider instead the
appropriate anisotropic expression, which can be written in
the form
\EQ
\alpha_{ip}=\tau\epsilon_{ijk}
\left(-\overline{u_ku_{j,p}}
+\rho_0^{-1}\overline{b_kb_{j,p}}\right).
\label{alpip}
\EN
The purpose of the present paper is to demonstrate that the values for both
expressions, \eq{alpiso} and \eq{alpip},
are almost identical in the cases presented by BS05.
We also show that the value of $\tau$, expressed in units of the turnover
time, is in all cases close to unity, and in some cases better so than in BS05.

\section{Formalism}

As in BS05 we consider cases where the flow is driven either by a
random body force in the momentum equation, or, alternatively, by random
externally imposed currents in the induction equation.
We calculated the isotropic expressions
\EQ
\tilde\alpha_{\rm K}=-\onethird\overline{\oo\cdot\uu},\quad
\tilde\alpha_{\rm M}=\onethird\rho_0^{-1}\overline{\jj\cdot\bb},
\label{alpisoKM}
\EN
in the presence of an imposed mean field $\meanBB_0$,
where the tilde indicates the absence of the $\tau$ factor, so
$\alpha=\tau(\tilde\alpha_{\rm K}+\tilde\alpha_{\rm M})$.
As in BS05, we use additional superscripts k and m to indicate
cases with kinetic or magnetic driving.
The resulting values of $\tilde\alpha_{\rm K}^{\rm(k)}$,
$\tilde\alpha_{\rm K}^{\rm(m)}$, $\tilde\alpha_{\rm M}^{\rm(k)}$,
and $\tilde\alpha_{\rm M}^{\rm(m)}$, presented below, are identical
to those of BS05.
In addition, we consider the appropriate component of the anisotropic
expressions for the same simulation data.
Since in our case the mean field points in the $y$ direction, and because
we use volume averages and periodic boundary conditions in all three
directions, we can write the anisotropic expressions for
$\tilde\alpha_{yy}$ in the form
\EQ
\tilde\alpha_{\rm K}^{\rm(a)}=-2\overline{u_xu_{z,y}},\quad
\tilde\alpha_{\rm M}^{\rm(a)}=2\rho_0^{-1}\overline{b_xb_{z,y}},
\label{alpipKM}
\EN
where the superscript (a) indicates anisotropy.
Again, we consider cases with kinetic and magnetic driving and
thus obtain the 4 values, $\tilde\alpha_{\rm K}^{\rm(ak)}$,
$\tilde\alpha_{\rm K}^{\rm(am)}$, $\tilde\alpha_{\rm M}^{\rm(ak)}$,
and $\tilde\alpha_{\rm M}^{\rm(am)}$.
The resulting values are normalized with respect to the
corresponding rms turbulent velocities,
\EQ
\tilde{a}^{\rm(k,m)}_{\rm K,M}=
\tilde\alpha^{\rm(k,m)}_{\rm K,M}/
\left[k_{\rm f}u_{\rm rms}u_{\rm rms}^{\rm(k,m)}\right],
\label{tildeadef}
\EN
where $u_{\rm rms}=[u_{\rm rms}^{\rm(k)}u_{\rm rms}^{\rm(m)}]^{1/2}$
is the geometrical mean of the rms velocities for kinetically and
magnetically driven runs.
This particular normalization emerges naturally when deriving the
time scale $\tau$ in \Eq{alpip}.
In the following we only consider the case of a statistically steady state,
so $\overline{b_xb_{z,y}}$ and $\overline{u_xu_{z,y}}$,
and hence also $\tilde\alpha_{\rm M}^{\rm(a)}$
and $\tilde\alpha_{\rm K}^{\rm(a)}$,
have converged to a stationary value.

\begin{figure}[t!]\begin{center}
\includegraphics[width=\columnwidth]{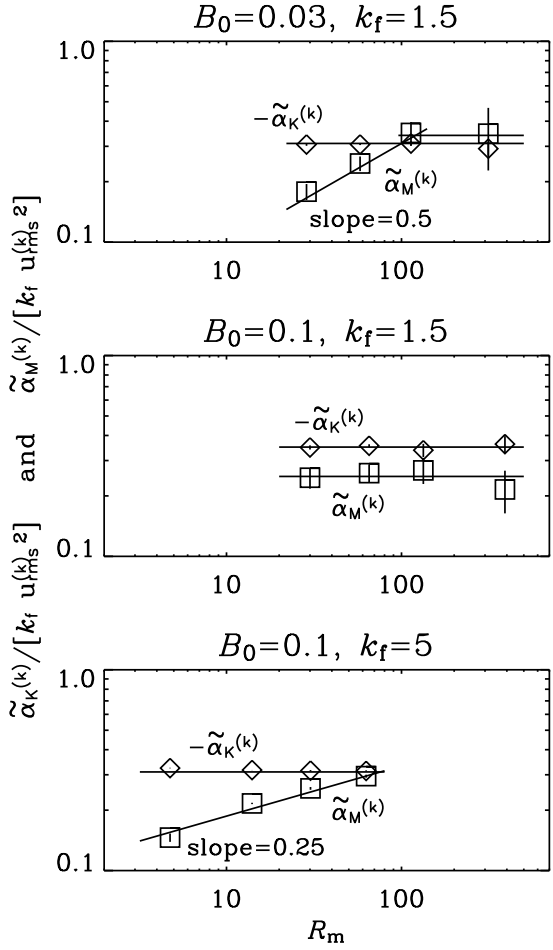}
\end{center}\caption[]{
Dependence of $\tilde{\alpha}_{\rm K}^{\rm(k)}$ and
$\tilde{\alpha}_{\rm M}^{\rm(k)}$
on $R_{\rm m}$ in the kinetically forced case.
Vertical bars give error estimates.
(Adapted from BS05.)
}\label{palpKMkin_vs_Rm}\end{figure}

\begin{figure}[t!]\begin{center}
\includegraphics[width=\columnwidth]{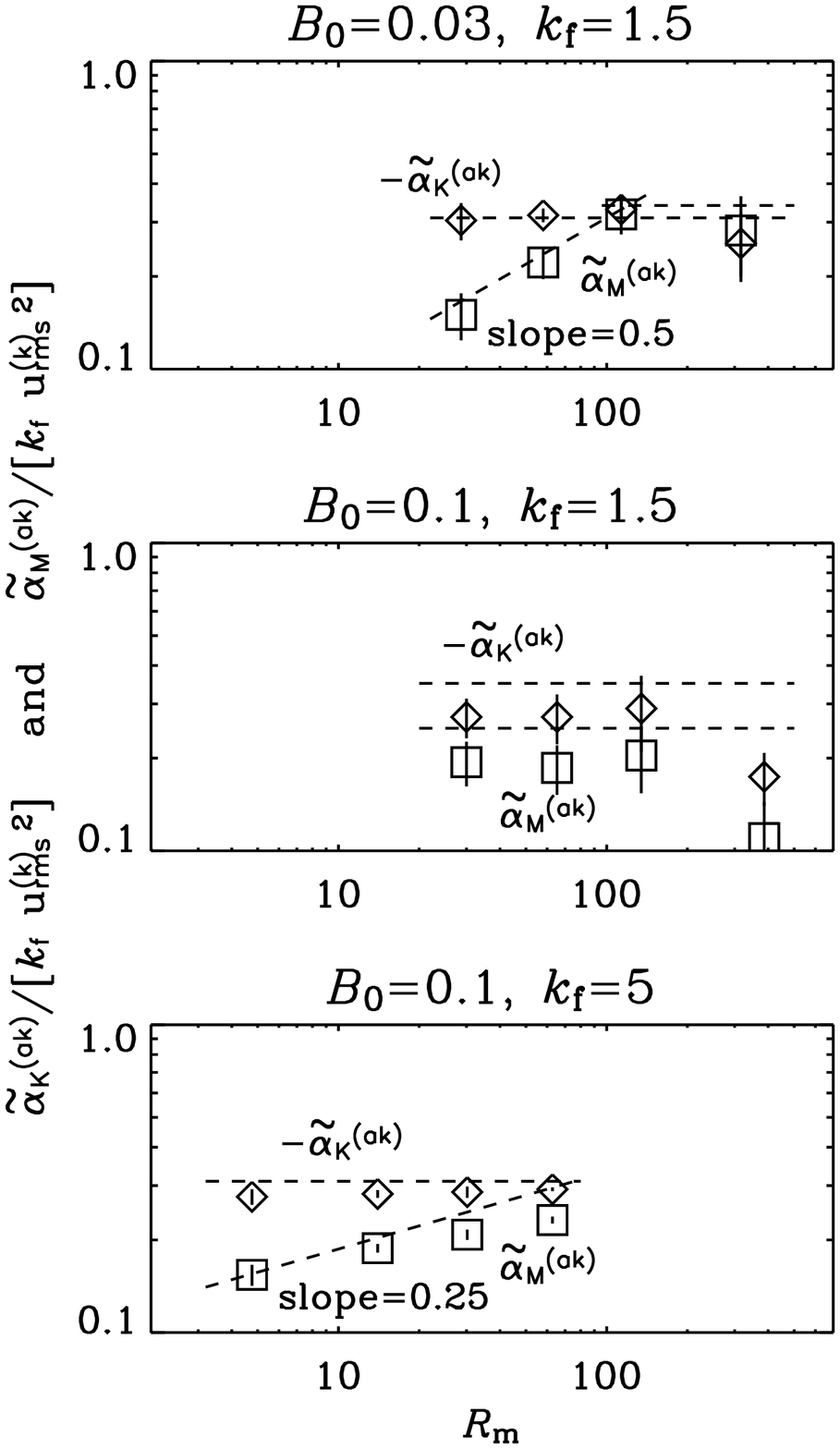}
\end{center}\caption[]{
Same as \Fig{palpKMkin_vs_Rm}, but for the relevant component of the
anisotropic expressions, $\tilde{\alpha}_{\rm K}^{\rm(ak)}$ and
$\tilde{\alpha}_{\rm M}^{\rm(ak)}$.
The dashed lines represent the fit to the data of \Fig{palpKMkin_vs_Rm},
not the present data!
}\label{paniKMkin_vs_Rm}\end{figure}

\section{Results}

\begin{figure}[t!]\begin{center}
\includegraphics[width=\columnwidth]{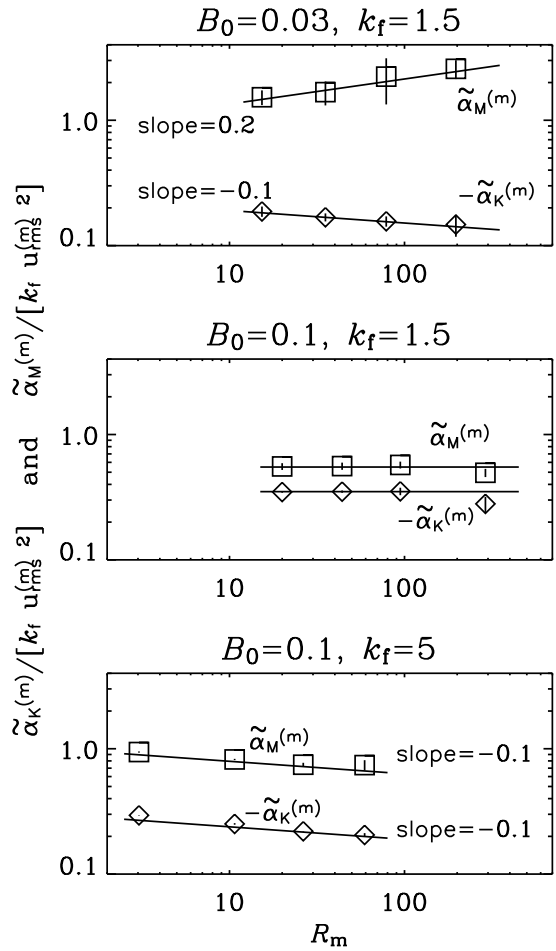}
\end{center}\caption[]{
Dependence of $\tilde{\alpha}_{\rm K}^{\rm(m)}$ and
$\tilde{\alpha}_{\rm M}^{\rm(m)}$
on $R_{\rm m}$ in the magnetically forced case.
Vertical bars give error estimates.
(Adapted from BS05.)
}\label{palpKMmag_vs_Rm}\end{figure}

We consider the values of $\tilde{\alpha}_{\rm K}$ and
$\tilde{\alpha}_{\rm M}$ and compare with the results of the appropriate
component of the
anisotropic expressions; see \Figs{palpKMkin_vs_Rm}{paniKMkin_vs_Rm} for
the kinetically driven case and \Figs{palpKMmag_vs_Rm}{paniKMmag_vs_Rm}
for the magnetically driven case.
The straight lines in \Figs{palpKMkin_vs_Rm}{palpKMmag_vs_Rm} denote
fits to the data points, while in \Figs{paniKMkin_vs_Rm}{paniKMmag_vs_Rm}
the same lines are just repeated as dashed lines
and still represent only the fits to
the isotropic data. This helps demonstrating that the results change
very little when the anisotropic expressions are used.

It is remarkable that the differences between the isotropic and
anisotropic expressions are rather systematic.
Generally speaking, the anisotropic expressions give
either the same or slightly smaller values than the isotropic expressions
if the flow is driven hydrodynamically.
The differences are larger for stronger fields ($B_0=0.1$) and especially
when the forcing it at larger scales ($k_{\rm f}$=1.5).
In that case the differences are around 15\% and 25\% for the
kinetic and magnetic $\alpha$ effects, respectively.
In the magnetically driven case the kinetic alpha effect tends
to be smaller for the anisotropic expressions, but the magnetic alpha
effect is either the same or larger for the anisotropic expressions.

Following BS05, we also compare the results for all runs in tabular form;
see \Tab{Ttau}.
As in BS05, we non-dimensionalize the measurements for
kinetically and magnetically driven cases independently, because
the root mean square velocities, $u_{\rm rms}^{\rm(k)}$ and
$u_{\rm rms}^{\rm(m)}$, are different in the two cases; see \Eq{tildeadef}.

\begin{figure}[t!]\begin{center}
\includegraphics[width=\columnwidth]{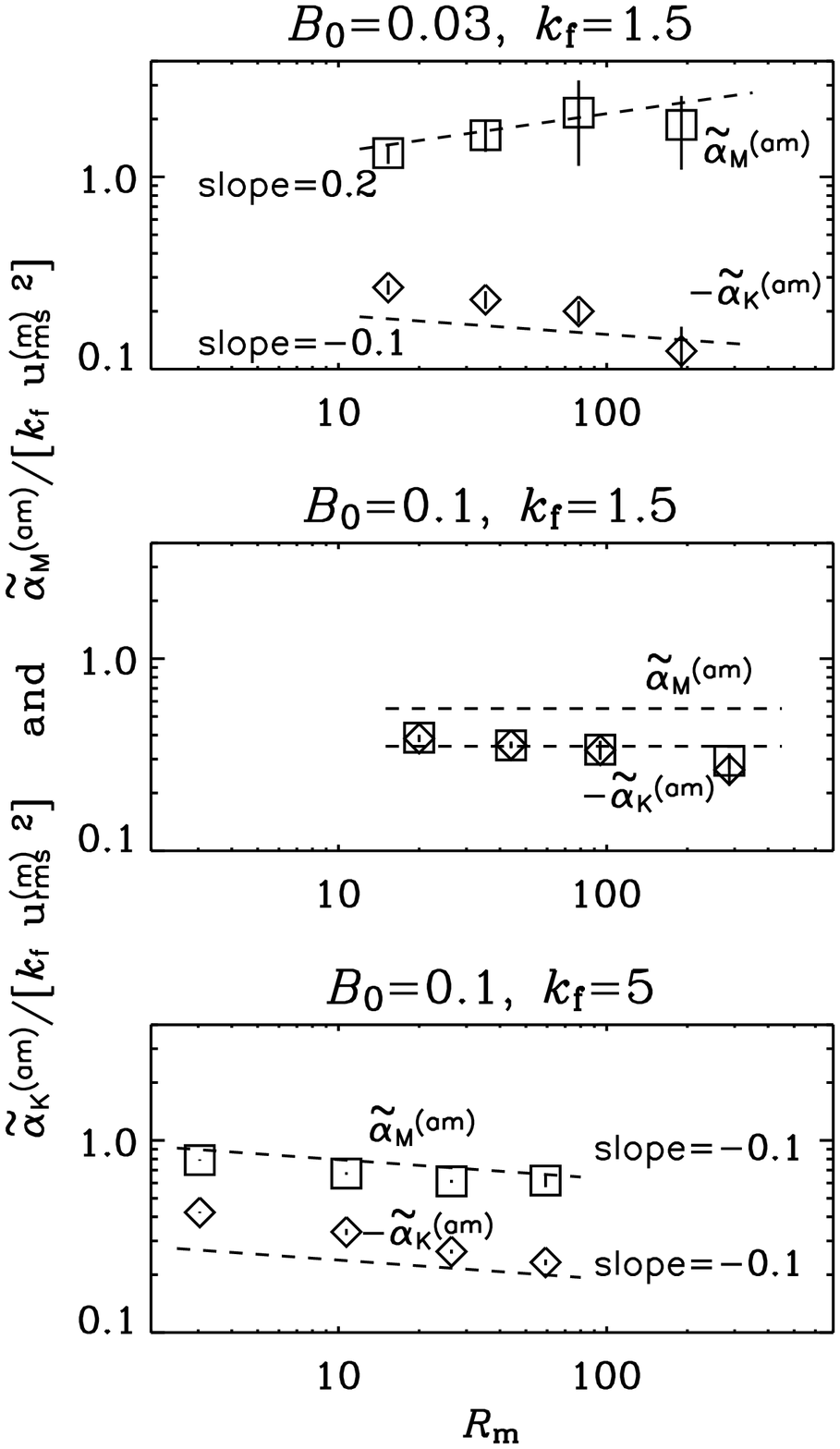}
\end{center}\caption[]{
Same as \Fig{palpKMmag_vs_Rm}, but for the relevant component of the
anisotropic expressions, $\tilde{\alpha}_{\rm K}^{\rm(am)}$ and
$\tilde{\alpha}_{\rm M}^{\rm(am)}$.
The dashed lines represent the fit to the data of \Fig{palpKMmag_vs_Rm},
not the present data!
}\label{paniKMmag_vs_Rm}\end{figure}

There are two important aspects of the $R_{\rm m}$ dependence of
kinetic and magnetic $\alpha$ effects.
One is the fact that, at least for moderate values of $R_{\rm m}$,
the two approach each other for finite
field strength and increasing strength of the mean field.
Furthermore, in the case of isotropic expressions, $|\tilde{\alpha}_{\rm M}|$
could even slightly exceed the value of $|\tilde{\alpha}_{\rm K}|$.
But when the anisotropic expressions are used, this is no longer the
case--or at least less drastically so, e.g.\ in the middle panel of
\Fig{paniKMkin_vs_Rm}.
The other aspect is the tendency for $\tilde{\alpha}_{\rm K}$ to stay
asymptotically independent of $R_{\rm m}$, even though the
actual $\alpha$ effect decreases like $1/R_{\rm m}^n$, with $n=0.5...1$,
as was shown in Fig.~2 of BS05 for the same data.
This property is critical to understanding the catastrophic
quenching of the $\alpha$ effect for closed or periodic
domains where magnetic helicity is a conserved quantity in
the high conductivity limit.
(We recall that, in contrast to the expressions for
$\tilde\alpha_{\rm K}^{\rm(a)}$ and $\tilde\alpha_{\rm M}^{\rm(a)}$,
$\alpha$ itself was always calculated as
$\alpha=\bra{\meanEMF\cdot\BB_0}_t/\BB_0^2$, which does already
account for the anisotropy for $\alpha$.
So the results for $\alpha$ remain unchanged from those obtained in BS05.)
Let us also note in this connection that, within error bars, the
off-diagonal components of the $\alpha$ tensor are found to be zero,
i.e.\ $|\bra{\meanEMF\times\BB_0}_t|=0$.

\begin{figure}[t!]\begin{center}
\includegraphics[width=\columnwidth]{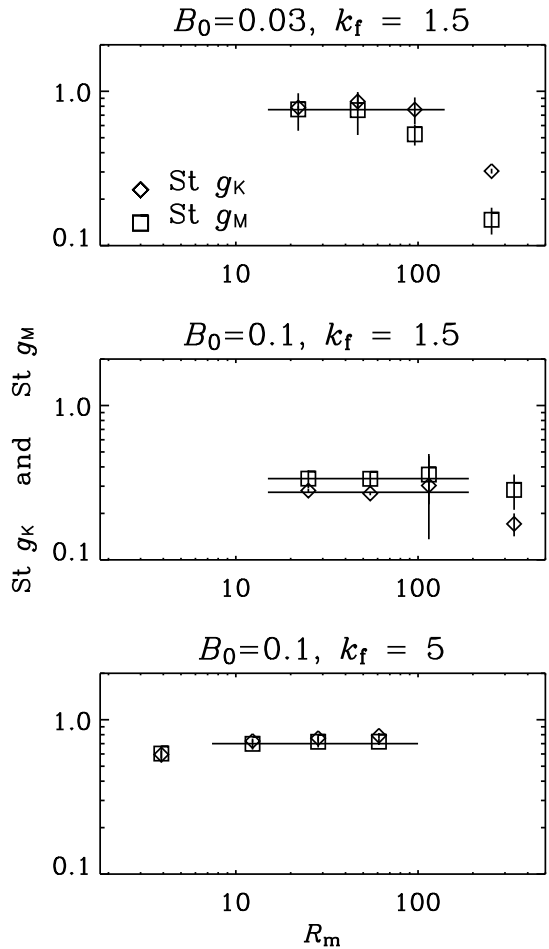}
\end{center}\caption[]{
Magnetic and kinetic Strouhal numbers as a function of $R_{\rm m}$
for different values of $B_0$ and  $k_{\rm f}$.
Here, kinetically and magnetically forced runs have been used to
calculate separately $g_{\rm K}\neq g_{\rm M}$.
The horizontal lines are drawn to indicate the range over which
the Strouhal numbers are approximately constant.
(Adapted from BS05.)
}\label{ptau_new}\end{figure}

\begin{figure}[t!]\begin{center}
\includegraphics[width=\columnwidth]{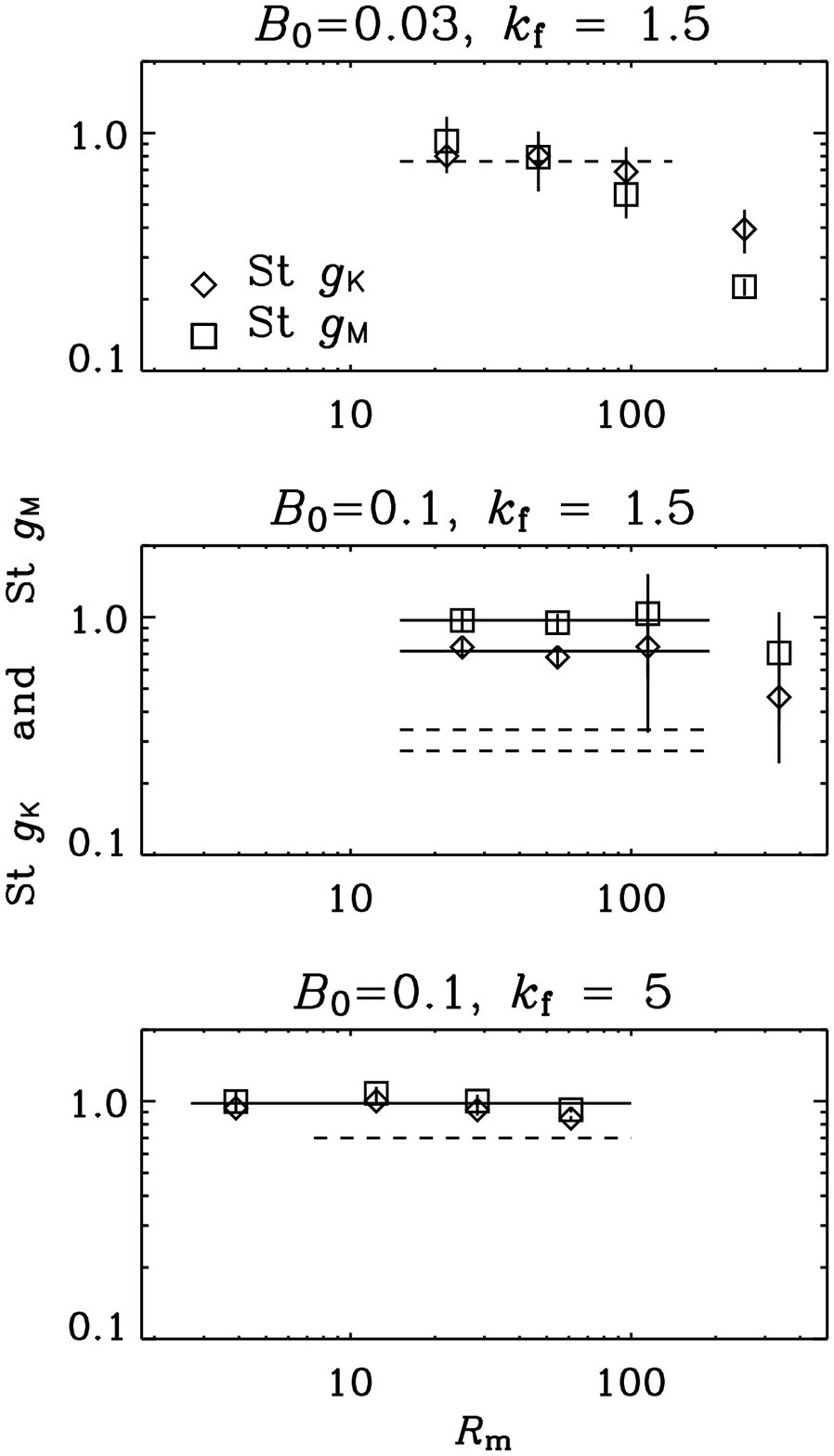}
\end{center}\caption[]{
Same as \Fig{ptau_new}, but for Strouhal numbers calculated
from the expressions for the anisotropic alpha coefficients.
The dashed lines represent the fits used in \Fig{ptau_new},
and the solid lines represent new fits.
}\label{ptau_aniso}\end{figure}

Finally we address the question of the relaxation time $\tau$.
In BS05 we calculated $\tau$ based on the values of $\alpha$,
$\tilde\alpha^{\rm(k,m)}_{\rm K}$, and $\tilde\alpha^{\rm(k,m)}_{\rm M}$.
In the following we repeat the same analysis using the anisotropic expressions,
$\tilde\alpha^{\rm(ak,am)}_{\rm K}$ and $\tilde\alpha^{\rm(ak,am)}_{\rm M}$.
We recall that we allowed for different and unknown prefactors
$g_{\rm K}$ and $g_{\rm M}$ in front of $\tilde\alpha_{\rm K}$ and
$\tilde\alpha_{\rm M}$.
We therefore wrote our unknowns in the form $\tau g_{\rm K}$ and
$\tau g_{\rm M}$, and expressed them in normalized form as
\EQ
\mbox{St}\,g_{\rm K,M}=u_{\rm rms}k_{\rm f}\tau g_{\rm K,M}.
\EN
These unknowns can be obtained by solving a matrix equation
which, in the present case, reads
\EQ
\pmatrix{a^{\rm(ak)}\cr a^{\rm(am)}}=
\pmatrix{
\tilde{a}_{\rm K}^{\rm(ak)}&\tilde{a}_{\rm M}^{\rm(ak)}\cr
\tilde{a}_{\rm K}^{\rm(am)}&\tilde{a}_{\rm M}^{\rm(am)}}
\pmatrix{\mbox{St}\,g_{\rm K}\cr\mbox{St}\,g_{\rm M}}.
\label{TwoUnknowns}
\EN
The result is shown in \Fig{ptau_new} for the old case using isotropic
expressions of $\tilde\alpha$, and in \Fig{ptau_aniso} for the present
case using the anisotropic expressions.

\begin{table*}[t]\caption{
Comparison of the results using the isotropic and anisotropic
expressions for the various values of the normalized $\alpha$
for kinetically and magnetically forced runs.
For $k_{\rm f}=1.5$ the resolution varies between $64^3$ and $512^3$
meshpoints for $\eta=2\times10^{-3}$ and $2\times10^{-4}$,
corresponding to magnetic Reynolds numbers of 20 and 300, respectively,
while for $k_{\rm f}=5$ the resolution varies between $32^3$ and $256^3$
meshpoints for $\eta=5\times10^{-3}$ and $5\times10^{-4}$,
corresponding to magnetic Reynolds numbers of 4 and 60, respectively,
The magnetic Prandtl number is always equal to unity, i.e.\ the
viscosity $\nu$ is always equal to the magnetic diffusivity, $\eta$.
}\vspace{12pt}\centerline{\begin{tabular}{lcc|cccccc|cccccc}
\hline
\hline
$B_0$ & $\eta$ & $k_{\rm f}$ &
$u_{\rm rms}^{\rm(k)}$ & $a^{\rm(k)}$ &
$\tilde{a}_{\rm K}^{\rm(k)}$ & $\tilde{a}_{\rm K}^{\rm(ak)}$ & $\tilde{a}_{\rm M}^{\rm(k)}$ & $\tilde{a}_{\rm M}^{\rm(ak)}$ &
$u_{\rm rms}^{\rm(m)}$ & $a^{\rm(m)}$ &
$\tilde{a}_{\rm K}^{\rm(m)}$ & $\tilde{a}_{\rm K}^{\rm(am)}$ & $\tilde{a}_{\rm M}^{\rm(m)}$ & $\tilde{a}_{\rm M}^{\rm(am)}$ \\
\hline
0.01&$2\times10^{-3}$&1.5&$ 0.10$&$-0.261$&$-0.46$&$-0.44$&$ 0.04$&$ 0.04$&$ 0.05$&$ 4.79$&$-0.11$&$-0.21$&$ 1.44$&$ 2.53$\\
0.03&$2\times10^{-4}$&1.5&$ 0.09$&$-0.048$&$-0.38$&$-0.33$&$ 0.46$&$ 0.36$&$ 0.06$&$ 0.29$&$-0.12$&$-0.10$&$ 2.23$&$ 1.44$\\
0.03&$5\times10^{-4}$&1.5&$ 0.09$&$-0.062$&$-0.37$&$-0.40$&$ 0.42$&$ 0.38$&$ 0.06$&$ 0.88$&$-0.13$&$-0.17$&$ 1.85$&$ 1.80$\\
0.03&$1\times10^{-3}$&1.5&$ 0.09$&$-0.099$&$-0.39$&$-0.40$&$ 0.32$&$ 0.28$&$ 0.05$&$ 0.88$&$-0.13$&$-0.18$&$ 1.31$&$ 1.29$\\
0.03&$2\times10^{-3}$&1.5&$ 0.09$&$-0.143$&$-0.42$&$-0.42$&$ 0.24$&$ 0.21$&$ 0.05$&$ 0.74$&$-0.14$&$-0.19$&$ 1.12$&$ 0.97$\\
0.06&$1\times10^{-3}$&1.5&$ 0.09$&$-0.030$&$-0.40$&$-0.39$&$ 0.36$&$ 0.28$&$ 0.06$&$ 0.23$&$-0.24$&$-0.28$&$ 0.61$&$ 0.46$\\
0.06&$2\times10^{-3}$&1.5&$ 0.08$&$-0.054$&$-0.40$&$-0.40$&$ 0.35$&$ 0.28$&$ 0.05$&$ 0.22$&$-0.24$&$-0.30$&$ 0.58$&$ 0.44$\\
0.10&$2\times10^{-4}$&1.5&$ 0.12$&$-0.003$&$-0.42$&$-0.20$&$ 0.24$&$ 0.13$&$ 0.09$&$ 0.07$&$-0.25$&$-0.23$&$ 0.41$&$ 0.25$\\
0.10&$5\times10^{-4}$&1.5&$ 0.10$&$-0.008$&$-0.41$&$-0.35$&$ 0.32$&$ 0.24$&$ 0.07$&$ 0.08$&$-0.29$&$-0.28$&$ 0.48$&$ 0.28$\\
0.10&$1\times10^{-3}$&1.5&$ 0.10$&$-0.010$&$-0.43$&$-0.33$&$ 0.32$&$ 0.23$&$ 0.07$&$ 0.08$&$-0.29$&$-0.29$&$ 0.46$&$ 0.29$\\
0.10&$2\times10^{-3}$&1.5&$ 0.09$&$-0.019$&$-0.43$&$-0.33$&$ 0.30$&$ 0.24$&$ 0.06$&$ 0.07$&$-0.28$&$-0.31$&$ 0.45$&$ 0.32$\\
0.14&$2\times10^{-3}$&1.5&$ 0.10$&$-0.009$&$-0.43$&$-0.25$&$ 0.26$&$ 0.20$&$ 0.06$&$ 0.04$&$-0.28$&$-0.28$&$ 0.45$&$ 0.26$\\
0.20&$2\times10^{-3}$&1.5&$ 0.11$&$-0.004$&$-0.43$&$-0.18$&$ 0.21$&$ 0.16$&$ 0.06$&$ 0.02$&$-0.27$&$-0.24$&$ 0.43$&$ 0.22$\\
0.30&$2\times10^{-3}$&1.5&$ 0.12$&$-0.002$&$-0.42$&$-0.14$&$ 0.18$&$ 0.13$&$ 0.06$&$ 0.01$&$-0.24$&$-0.19$&$ 0.41$&$ 0.19$\\
0.06&$5\times10^{-4}$&5  &$ 0.16$&$-0.080$&$-0.31$&$-0.30$&$ 0.25$&$ 0.22$&$ 0.15$&$ 0.08$&$-0.25$&$-0.20$&$ 1.10$&$ 0.45$\\
0.06&$1\times10^{-3}$&5  &$ 0.16$&$-0.121$&$-0.32$&$-0.30$&$ 0.20$&$ 0.18$&$ 0.14$&$ 0.01$&$-0.12$&$-0.09$&$ 2.03$&$ 0.17$\\
0.06&$2\times10^{-3}$&5  &$ 0.15$&$-0.172$&$-0.49$&$-0.46$&$ 0.22$&$ 0.20$&$ 0.06$&$ 0.34$&$-0.16$&$-0.22$&$ 0.52$&$ 0.44$\\
0.06&$5\times10^{-3}$&5  &$ 0.13$&$-0.215$&$-0.41$&$-0.37$&$ 0.10$&$ 0.11$&$ 0.08$&$ 0.54$&$-0.18$&$-0.23$&$ 0.81$&$ 0.72$\\
0.10&$5\times10^{-4}$&5  &$ 0.16$&$-0.035$&$-0.32$&$-0.30$&$ 0.30$&$ 0.24$&$ 0.15$&$ 0.36$&$-0.20$&$-0.23$&$ 0.72$&$ 0.60$\\
0.10&$1\times10^{-3}$&5  &$ 0.15$&$-0.058$&$-0.34$&$-0.31$&$ 0.27$&$ 0.22$&$ 0.13$&$ 0.35$&$-0.21$&$-0.25$&$ 0.70$&$ 0.57$\\
0.10&$2\times10^{-3}$&5  &$ 0.14$&$-0.091$&$-0.36$&$-0.32$&$ 0.25$&$ 0.22$&$ 0.11$&$ 0.34$&$-0.22$&$-0.29$&$ 0.72$&$ 0.59$\\
0.10&$5\times10^{-3}$&5  &$ 0.12$&$-0.131$&$-0.41$&$-0.35$&$ 0.18$&$ 0.19$&$ 0.08$&$ 0.31$&$-0.24$&$-0.34$&$ 0.75$&$ 0.63$\\
\hline
\label{Ttau}\end{tabular}}\end{table*}

One of the most remarkable results from \Fig{ptau_aniso} is that 
the values of the magnetic and kinetic Strouhal numbers are in all three
cases close to unity, whereas in the middle panel of \Fig{ptau_new} the
Strouhal numbers were only about 0.3.
In all other aspects the new results are rather similar to the old ones.
For example, the values of magnetic and kinetic Strouhal numbers are
rather close to each other except in the case $B_0=0.1$ with $k_{\rm f}=1.5$,
where the magnetic Strouhal numbers are somewhat larger than the kinetic ones.
This is also the parameter regime for which the largest differences
were found between \Figs{palpKMkin_vs_Rm}{paniKMkin_vs_Rm}.
Furthermore, like in BS05, we still find a drop in the Strouhal numbers
in the case where $R_{\rm m}$ is around 300.
As argued in BS05, this may be connected with these simulations not having
run for long enough.

\section{Discussion}

The work of BS05 was mainly an extension of earlier work on
passive scale diffusion (Brandenburg et al.\ 2004), where certain aspects
of MTA were tested.
In particular, it was shown that the relaxation time $\tau$ in the
$\tau$ approximation is of the order of the turnover time
($\mbox{St}=\tau u_{\rm rms}k_{\rm f}\approx3$).
In the case with a magnetic field, the $\alpha$ effect was assumed to be
expressible as $\alpha=\tau(\tilde{\alpha}_{\rm K}+\tilde{\alpha}_{\rm M})$.
The main result of BS05 was that $\mbox{St}$ is independent of $R_{\rm m}$.
This is important because neither $\tilde{\alpha}_{\rm K}$ nor
$\tilde{\alpha}_{\rm M}$ decline with increasing values of $R_{\rm m}$.
Instead, $-\tilde{\alpha}_{\rm M}$ approaches $\tilde{\alpha}_{\rm K}$,
resulting in near cancellation.
Together with the finding that $\tau$ is approximately independent
of $R_{\rm m}$, this supports the validity of the assumed formula
for $\alpha$.
It should be noted, however, that for $R_{\rm m}\approx300$ the result is
not convincing and our present data suggest a drop in the Strouhal number.

However, as RR07 have pointed out, several other issues remained open
or unsatisfactory.
In particular the comparative use of kinetically and magnetically
forced models may be questionable.
This was done to change the relative importance of kinetic and magnetic
$\alpha$ effects.
The problem is that the nature of the turbulence can change considerably
in the two cases.
On the other hand, there is no reason why the expressions for $\alpha$
should not apply equally well in both regimes

Another problem is the use of isotropic expressions for
$\tilde{\alpha}_{\rm K}$ and $\tilde{\alpha}_{\rm M}$.
Surprisingly enough, as we have shown here, the isotropic expressions
are indeed good pro\-xies for the relevant component of the
full anisotropic expressions.
One advantage of using the anisotropic expressions is that the need for
adopting (slightly) different coefficients in front of
$\tilde{\alpha}_{\rm K}$ and $\tilde{\alpha}_{\rm M}$ is now less
severe, if at all present.

Finally, there is the puzzle that, on the one hand,
when using the first order smoothing approximation (FOSA), $\alpha$ is given
by an expression involving just the actual velocity field while, on the
other hand, according to the $\tau$ approximation, it is the sum of
magnetic and kinetic $\alpha$ effects.
Obviously, a rigorous comparison between FOSA and $\tau$ approximation
is only permissible when the magnetic Reynolds number is below unity.
In the present paper this is not the case, so the neglect of the higher
order (triple) correlation terms under FOSA cannot be justified, given
that the Strouhal numbers are always around unity.
So this comparison may not have been permissible.
However, the puzzle seems to exist even in the low magnetic Reynolds
number limit, when the triple correlations can be neglected altogether.
This case has been analyzed recently by Sur et al.\ (2007), who showed
that the formulations in terms of FOSA and $\tau$ approximation
are in fact equivalent (as they have to be, because the starting equations
are the same!), but that the individual components contributing to
the total $\alpha$-effect in the two formulations are different.
In fact, it turns out that in the $\tau$ approximation there is,
in addition to the kinetic and magnetic alpha effects,  
in general also one more term resulting from the correlation
between the small scale magnetic field and the forcing function.
Only in the special case of $\delta$-correlated forcing, 
that is adopted in many numerical investigations, does this extra
term vanish.
Nevertheless, even then the kinetic part of the alpha effect in the
$\tau$ approximation is not simply related to the alpha effect obtained
from the first order smoothing approximation, even if the actual velocity
field is used in both cases.
Therefore there is actually no puzzle in the limit of small magnetic
Reynolds numbers either.

\section{Conclusions}

We have shown that the basic conclusions obtained in BS05 carry over to
the case where the anisotropic expressions for $\tilde{\alpha}_{\rm K}$
and $\tilde{\alpha}_{\rm M}$ are used.
The present work provides an extra piece of evidence that
the $\tau$ approximation may provide a useable formalism for describing
simulation data and for predicting the behavior in situations that are
not yet accessible to direct simulations.
There are currently no other approaches capable of this.
The basic hypothesis that the triple correlations are expressible as a
damping term may not be well justified, although some important properties
of this approach seem to be borne out by simulations.
A number of further practical tests of the $\tau$ approximations could
be envisaged.
One such example might be the so-called $\meanWW\times\meanJJ$ effect
of Rogachevskii \& Kleeorin (2003, 2004),
which was derived using the $\tau$ approximation.
Direct simulations of hydromagnetic turbulence with shear give qualitative
support to this idea (Brandenburg 2005a), although it is not clear
under which conditions the anticipated effect has the appropriate sign
for dynamo action (Brandenburg 2005b; R\"udiger \& Kitchatinov 2006;
R\"adler \& Stepanov 2006).
Further work in this direction would be worthwhile for establishing
the real usefulness of the $\tau$ approximation.

\acknowledgements
The Danish Center for Scientific Computing and the Center for Parallel
Computers at KTH are acknowledged for providing computer time.


\end{document}